\documentclass{aastex}
\usepackage{emulateapj5}
\usepackage{times,mathptmx}

\slugcomment{Submitted to ApJ}

\shorttitle{A comparison of stellar mass estimates}
\shortauthors{Drory, Bender, \& Hopp}

\newcommand*{\Msun}{\ensuremath{\mathrm{M_\odot}}}%
\newcommand*{\MLK}{\ensuremath{M/L_K}}%
\newcommand*{\MLg}{\ensuremath{M/L_g}}%
\newcommand*{\ML}{\ensuremath{M/L}}%
\newcommand*{\Mdyn}{\ensuremath{M_{\mathrm{dyn}}}}%

\newcommand*{\Ha}{H$\alpha$}%
\newcommand*{\Hd}{H$\delta_{\mathrm{A}}$}%
\newcommand*{\D}{D$_\mathrm{n}$4000}%

\begin{document}

%% ------------------------------------------------------------------
%% TITLE
%% ------------------------------------------------------------------

\title{Comparing spectroscopic and photometric stellar mass estimates}

%% ------------------------------------------------------------------
%% AUTHORS
%% ------------------------------------------------------------------

\author{N.~Drory\altaffilmark{1},
        R.~Bender\altaffilmark{2,3}, U.~Hopp\altaffilmark{3}}

\affil{$^1$ University of Texas at Austin, Austin, Texas 78712}
\email{drory@astro.as.utexas.edu}

\affil{$^2$ Max--Planck Institut f\"ur extraterrestrische Physik,
  Giessenbachstra\ss e, Garching, Germany}
\email{bender@mpe.mpg.de}

\affil{$^3$ Universit\"ats--Sternwarte M\"unchen, Scheinerstra\ss
  e 1, D-81679 M\"unchen, Germany}
\email{hopp@usm.uni-muenchen.de}

%% ------------------------------------------------------------------
%% ABSTRACT
%% ------------------------------------------------------------------

\begin{abstract}
  The purpose of this letter is to check the quality of different
  methods for estimating stellar masses of galaxies. We compare the
  results of (a) fitting stellar population synthesis models to broad
  band colors from SDSS and 2MASS, (b) the analysis of spectroscopic
  features of SDSS galaxies \citep{Kauffmannetal03a}, and, (c) a
  simple dynamical mass estimate based on SDSS velocity dispersions
  and effective radii.  Knowing that all three methods can have
  significant biases, a comparison can help to establish their
  (relative) reliability. In this way, one can also probe the quality
  of the observationally cheap broadband color mass estimators for
  galaxies at higher redshift.  Generally, masses based on broad-band
  colors and spectroscopic features agree reasonably well, with a rms
  scatter of only $\sim 0.2$~dex over almost 4 decades in mass.
  However, as may be expected, systematic differences do exist and
  have an amplitude of $\sim 0.15$~dex, corrleting with \Ha\ emission
  strength.  Interestingly, masses from broad-band color fitting are
  in better agreement with dynamical masses than masses based on the
  analysis of spectroscopic features. In addition, the differences
  between the latter and the dynamical masses correlate with \Ha\ 
  equivalent width, while this much less the case for the broad-band
  masses.  We conclude that broad band color mass estimators, provided
  they are based on a large enough wavelength coverage and use an
  appropriate range of ages, metallicities and dust extinctions, can
  yield fairly reliable stellar masses for galaxies. This is a very
  encouraging result as such mass estimates are very likely the only
  ones available at significant redshifts for some time to come.
\end{abstract}

%% KEYWORDS
%%
\keywords{galaxies: mass function --- galaxies: fundamental parameters}

%% ------------------------------------------------------------------
%% INTRODUCTION
%% ------------------------------------------------------------------

\section{Introduction}\label{sec:introduction}

The stellar mass of galaxies at the present epoch and the build-up of
stellar mass over cosmic time has become the focus of intense research
in the past few years.

In the local universe, results on the stellar mass function of
galaxies were published using the new generation of wide-angle surveys
in the optical (Sloan Digital Sky Survey; SDSS, \citealp{SDSS}; 2dF,
e.g.~\citealp{2dF99}) and near-infrared (Two Micron All Sky Survey;
2MASS, \citealp{TwoMASS}). \citet{2dF01} combined data from 2MASS and
2dF to derive the local stellar mass function, \citet{BMKW03} used the
SDSS and 2MASS to the same end.

At $z > 0$, a number of authors studied the stellar mass density as a
function of redshift
\citep{BE00,MUNICS3,Cohen02,DPFB03,Fontanaetal03,Rudnicketal03}
reaching $z \sim 3$, while others, using wider field surveys,
investigated the evolution of the mass function of galaxies
\citep{MUNICS6,K20-04} to $z \sim 1.5$.

Generally, the high-redshift work relies on fits of multi-color
photometry to a grid of composite stellar population (CSP) models to
determine a stellar mass-to-light ratio, since large and complete
spectroscopic samples of galaxies are not yet available.  A similar
approach was chosen by \citet{2dF01} and \citet{BMKW03}, too, at $z
\sim 0$.

Taking advantage of the availability of photometry and spectroscopy
for galaxies in the SDSS, \citet[][K03 hereafter]{Kauffmannetal03a}
utilized spectroscopic diagnostics (4000\AA~Break, \D, and the \Hd\ 
Balmer absorption line index ) to estimate the mean stellar age and
the fraction of stars formed in recent bursts in each galaxy.  By
comparison of the colors predicted by their best-fit model to the
object's broad-band photometry they determine the amount of extinction
by dust and hence the stellar mass-to-light ratio.

The purpose of this letter is to compare the stellar masses determined
by this spectroscopic technique to masses obtained from multi-passband
photometry and to compare both methods to a simple dynamical estimate
of mass. Knowing that none of these methods yields a fiducial
(stellar) mass, a comparison helps to establish the (relative)
reliability of each method and makes us aware of potential differences
between these estimators. Moreover, it can show us whether one can use
observationally cheaper estimators as surrogates for more expensive
(or unobtainable) ones, which is particularly important when dealing
with high-redshift datasets.

Specifically, we want to know how the two estimators compare to each
other, if using K-band \ML\ yields better masses than using the g-band
\ML\ which is accessible at high $z$, and how these compare to a simple
dynamical mass estimator, $M \sim \sigma^2 R_e / G$.

\begin{figure*}[t]
  \centering
  \epsscale{0.8}
  %\epsscale{0.45}
  \plotone{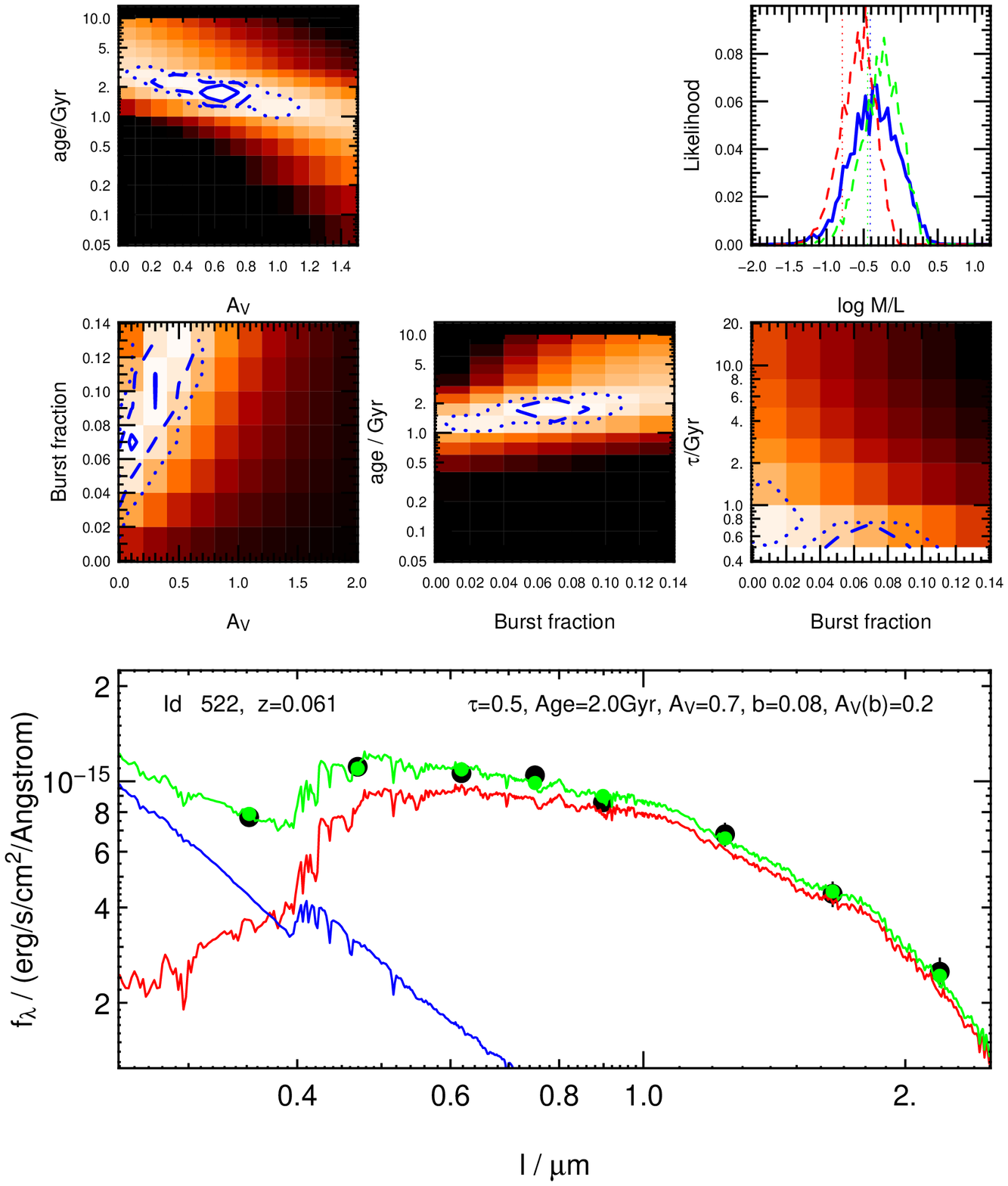}\hspace*{0.5cm}
  \plotone{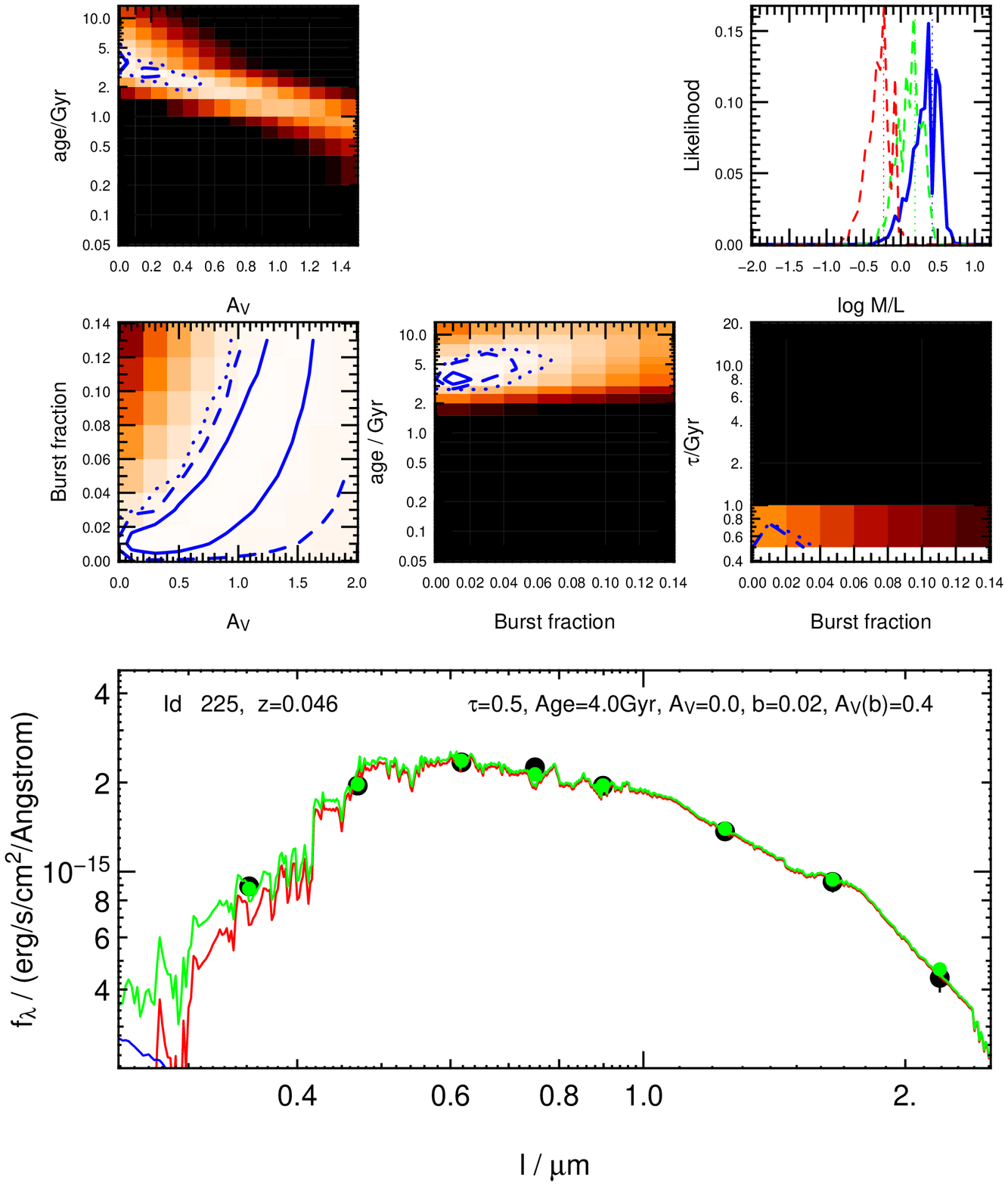}
  \caption{\label{fig:loglik}%
    Illustration of the model-fitting technique used to estimate
    stellar masses. The lower panels show the comparison of the best
    fitting model to the photometric data. The red and blue lines
    represent the main and burst component, respectively. The green
    line represents the combined SED. The left hand side shows a young
    object with a burst component, the right side an older object.
    The upper panels show projections of the likelihood function onto
    four planes, age vs.\ dust in the main component, burst fraction
    vs.\ burst extinction, age of the main component vs.\ burst
    fraction, and star formation timescale, $\tau$, vs.\ burst
    fraction. The resulting likelihood distributions of \ML\ are shown
    in the upper right panel on each side ($M/L_g$ blue; $M/L_i$
    green; $M/L_K$ red). The \ML\ of the best fitting model is
    indicated by vertical lines.}
\end{figure*}

This letter is laid out as follows. In Sect.~\ref{sec:galaxy-sample}
we describe the sample of galaxies we use in this work.  In
Sect.~\ref{sec:deriving-masses} we give a brief overview of how we
derive stellar masses by fitting CSP models to multi-band photometry.
In Sect.~\ref{sec:stellmass} we compare these masses to the values in
K03 and in Sect.~\ref{sec:dynmass} to a simple dynamical estimate of
mass based on the SDSS velocity dispersions. We also discuss the
implications of these comparisons.

We assume $\Omega_{\mathrm{M}} = 0.3$, $\Omega_{\Lambda} = 0.7$, and
$H_0 = 70~\mathrm{km\ s^{-1}\ Mpc^{-1}}$ throughout this work.

%% ------------------------------------------------------------------
%% THE GALAXY SAMPLE
%% ------------------------------------------------------------------

\section{The galaxy sample}\label{sec:galaxy-sample}

The sample of galaxies we use in this work is selected from the NYU
Value-Added Galaxy Catalog\footnote{see also
  \texttt{http://wassup.physics.nyu.edu/vagc/}} \citep{VAGC}. This is
a merged catalog of objects from the SDSS Data Release Two (DR2) and
2MASS point-source and extended-source catalogs (and other catalogs
which are not relevant here, as well).

We select all objects classified as galaxies and having a secure
redshift measurement in the SDSS and that are detected in the 2MASS
catalogs. From this set we randomly sub-select 20\% of the objects
leaving us with a sample of sample of $\sim 17000$ objects having
redshifts and photometry in ugrizJHK.

We cross-correlate this catalog with the data from
K03\footnote{available online at
  \texttt{http://www.mpa-garching.mpg.de/SDSS/}} to obtain their
stellar mass estimates.

The galaxies in the sample span the absolute magnitude range $-15.3 <
M_g < -23.5$, the restframe $u\!-\!g$ color range $0.5 < u\!-\!g <
2.0$, and the (stellar) mass range $8 < \log M < 12$.

%% ------------------------------------------------------------------
 %% DERIVING STELLAR MASSES
%% ------------------------------------------------------------------

\section{Deriving stellar masses}\label{sec:deriving-masses}

\begin{figure*}[t]
  \centering
  \epsscale{0.80}
  %\epsscale{0.40}
  \plotone{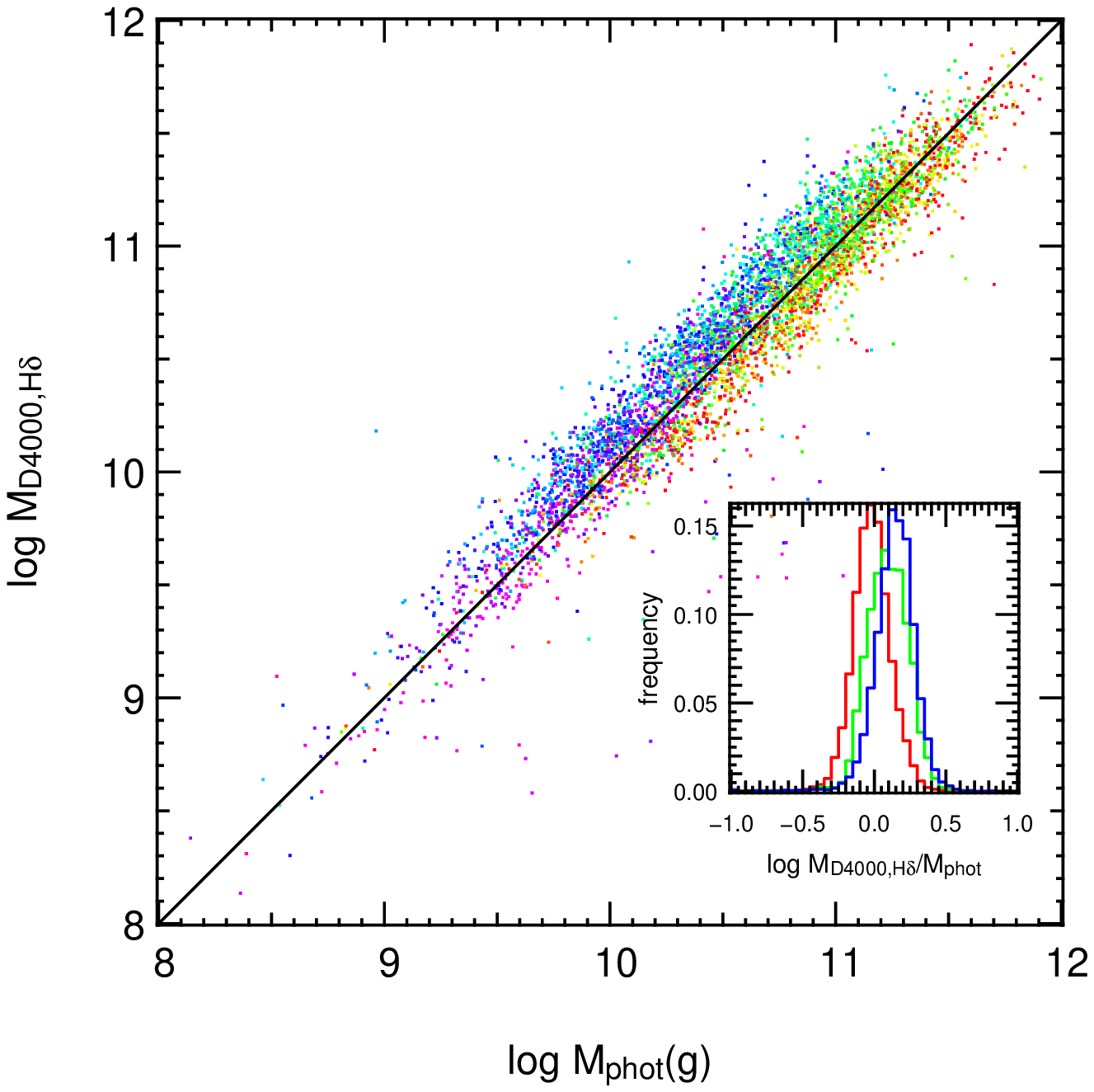}\hspace*{1cm}
  \plotone{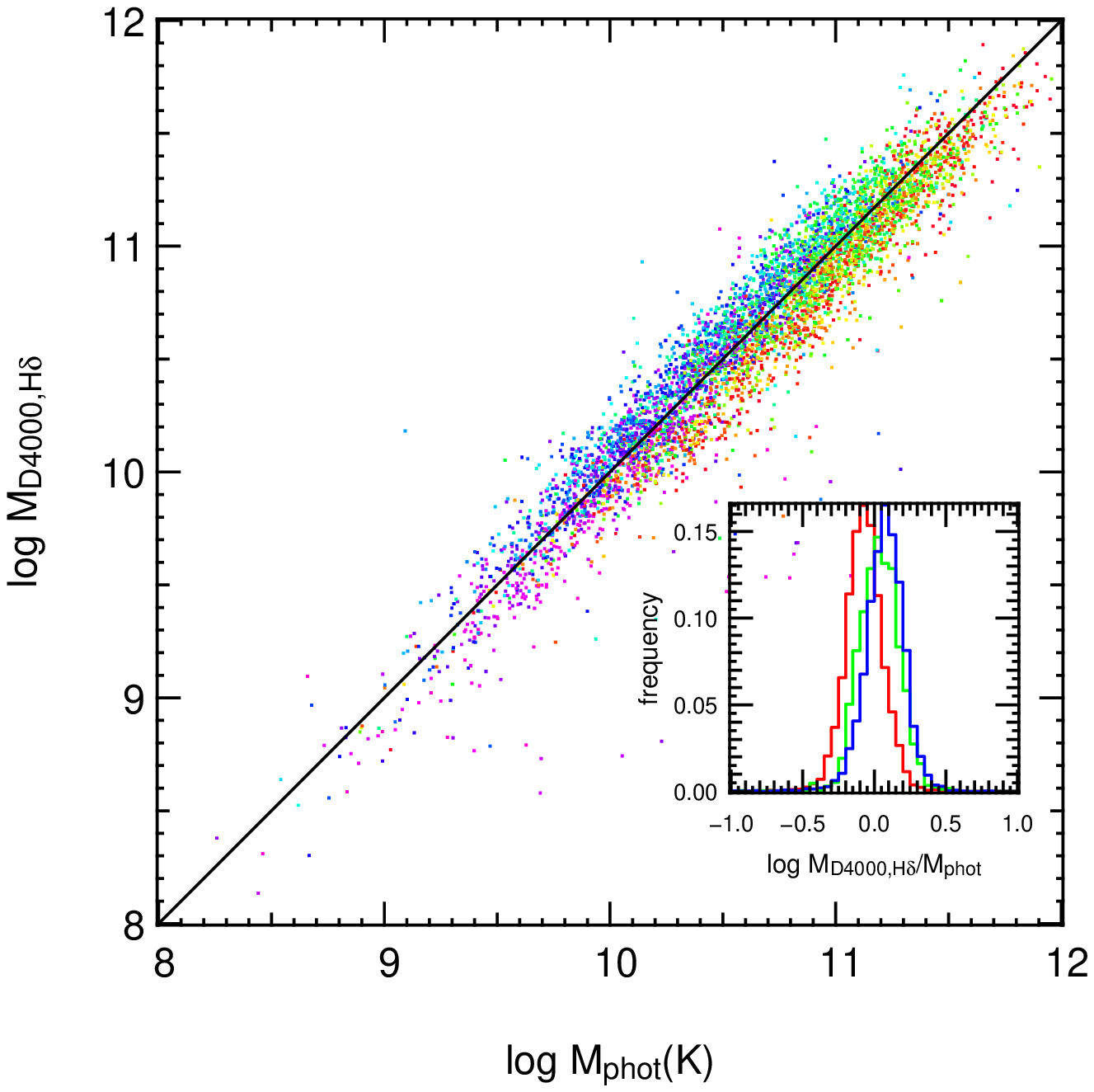}
  \caption{\label{fig:mass-comp}%
    Comparison of our photometry-based stellar mass estimates to the
    stellar masses of K03. The left panel shows the masses of K03
    plotted against our estimate based on $M/L_g$, the right hand
    panel against masses based on $M/L_K$. The colors denote \Ha\ 
    equivalent width from no emission (red) to strong emission ($>
    25$\AA; purple). The small panels show histograms of the residuals
    again as a function of \Ha\ equivalent width.}
\end{figure*}

The method we use to infer stellar masses from multi-color photometry
is an advancement of the program used in \citet{MUNICS6}. It is based
on the comparison of multi-color photometry to a grid of stellar
population synthesis models covering a wide range in parameters,
especially star formation histories (SFHs).

We base our new model grid on the \citet{BC03} stellar population
synthesis package.  We parameterize the possible SFHs by a
two-component model, consisting of a main component with a smooth
analytically described SFH and a burst of star formation.  The main
component is parameterized by a star formation rate of the form
$\psi(t) \propto \exp(-t/\tau)$, with $\tau \in [0.1, \infty]$~Gyr and
a metallicity of $-0.6 < \mathrm{[Fe/H]} < 0.3$.  The age, $t$, is
allowed to vary between 0.5~Gyr and the age of the universe (at the
object's redshift).

The smooth component is linearly combined with a burst of star
formation, which is modeled as a 100~Myr old constant star formation
rate episode of solar metallicity. We restrict the burst fraction,
$\beta$, to the range $0 < \beta < 0.15$ in mass (higher values of
$\beta$ are degenerate and unnecessary since this case is covered by
models with a young main component).  We adopt a Salpeter initial mass
function for both components, with lower and upper mass cutoffs of 0.1
and 100~\Msun.

Additionally, both the main component and the burst are allowed to
exhibit a variable amount of extinction by dust.  This takes into
account the fact that young stars are found in dusty environments and
that the starlight from the galaxy as a whole may be reddened by a
(geometry dependent) different amount. In fact, \citet{SMSS04} find
that the extinction derived from the Balmer decrement in the SDSS
sample is independent of inclination, which, on the other hand, is
driving global extinction (\citealp[see, e.g.,][]{TPHSVW98}). This is
different from the approach taken by K03, where a single extinction
value for the whole galaxy is used.

We compute the full likelihood distribution on a grid in this
6-dimensional parameter space ($\tau, \mathrm{[Fe/H]}, t, A_V^1,
\beta, A_V^2$), the likelihood of each model being $\propto
\exp(-\chi^2/2)$. To compute the likelihood distribution of \ML, we
weight the \ML\ of each model by its likelihood and marginalize over
all parameters. The uncertainty in \ML\ is obtained from the width of
this distribution.

This procedure is illustrated in Fig.~\ref{fig:loglik}, where we show
SEDs and likelihood functions for two objects, a young object with a
high burst fraction, and an older and fairly quiescent object. We show
projections of the likelihood function onto four planes in parameter
space, age vs.\ dust in the main component, burst fraction vs.\ burst
extinction, age of the main component vs.\ burst fraction, and star
formation timescale, $\tau$, vs.\ burst fraction.  The figure also
shows the resulting likelihood distributions of \ML\ in the g, i, and
K bands. Note that for the quiescent object, the width of the \ML\ 
distribution is very similar in the g and K bands, while it is much
wider in g than it is in K for the younger star forming object.  On
average, the width of the likelihood distribution of \ML\ at 68\%
confidence level is between $\pm 0.1$ and $\pm 0.2$~dex (using \MLg).
The uncertainty in mass has a weak dependence on mass (increasing with
lower $S/N$ photometry) and much of the variation is in spectral type:
early-type galaxies have more tightly constrained masses than late
types (see also Fig.~\ref{fig:loglik}).  Using the U band, the
uncertainty in mass grows by $\sim 0.05$~dex.

%% ------------------------------------------------------------------
%% COMPARISON OF STELLAR MASSES
%% ------------------------------------------------------------------

\section{Comparison of stellar mass estimators}\label{sec:stellmass}

In Fig.~\ref{fig:mass-comp} we compare our photometry-based stellar
mass estimates to the stellar masses of K03. We show the \Ha\ 
equivalent width (as measured by the SDSS) by color coding. The
overall impression from this figure is that the two different
estimators agree remarkably well, within a rms scatter of only $\sim
0.2$ dex over almost 4 decades in mass (and hence they largely agree
within their respective uncertainties). This is only a relative
statement, though. It does not imply that the masses are accurate to
that level in an absolute sense, although it is very reaffirming.
However, as may be expected, there are systematic differences as a
function of star formation activity on the $\sim 0.15$~dex level.

The \D\ and \Hd\ based method of K03 yields masses almost identical to
ours for weakly star forming objects (EqW(\Ha) $< 5$\AA) at masses
above $10^{10.5}$~\Msun. At lower masses, our estimator tends to give
slightly higher masses than K03's. For more strongly star forming
objects, the photometrically determined masses are smaller than the
ones of K03. For objects with EqW(\Ha) $> 25$\AA, the discrepancy
becomes as large as 0.15 dex, independent of mass. Note that at high
redshift, such objects will be more common.

We suspect that there are a multitude of reasons for these differences
based on the different sampling of stellar populations by both methods
(if we leave out the JHK bands, our masses become more similar in
their trends to K03's, although with increased scatter). Plausibly,
though, this is explainable by the fact that the photometrically
determined masses sample the light from the whole galaxy, while K03's
SDSS-based sampling of \D\ and \Hd\ covers only the inner 3
arcseconds. Since most galaxies are redder in their centers than in
the outer parts, this might lead to higher masses for star forming
disk galaxies.  Early-type galaxies without blue star forming disks do
not suffer from this effect. This is confirmed by restricting the
sample to low redshifts, which maximizes the effect.  Also, at the
lowest masses, galaxies might have more irregular SFHs, and
photometric methods might fail in this case \citep{BD01}. However,
Fig.\ref{fig:mass-comp} does not show a dependence of the residuals on
mass, only on current star formation rate.

Fig.~\ref{fig:mass-comp} also shows that our masses based on the g
band are very similar to the ones estimated through the K band. For
early type systems and weakly star forming systems, they are
statistically indistinguishable. Star forming systems, however, tend
to have g-band masses lower by $\sim 0.1$~dex. The good agreement is
partly due to the fact that the effect of dust extinction and age on
the effective \ML\ are very similar in canonical models, as has been
pointed out by \citet{BD01}, although they are not completely
degenerate. Typically, the spread of the stellar \ML\ in the galaxy
population at any given luminosity is around 0.7~dex in g and 0.35~dex
in K. These results is reaffirming since the restframe blue spectral
range is accessible to photometry to very high redshift, and thus
high-$z$ studies mostly rely on $M/L_B$.

%% ------------------------------------------------------------------
%% DYNAMICAL MASSES
%% ------------------------------------------------------------------

\section{Comparison with dynamical masses}\label{sec:dynmass}

\begin{figure*}[t]
  \centering
  %\epsscale{0.7}
  \epsscale{1.3}
  \plotone{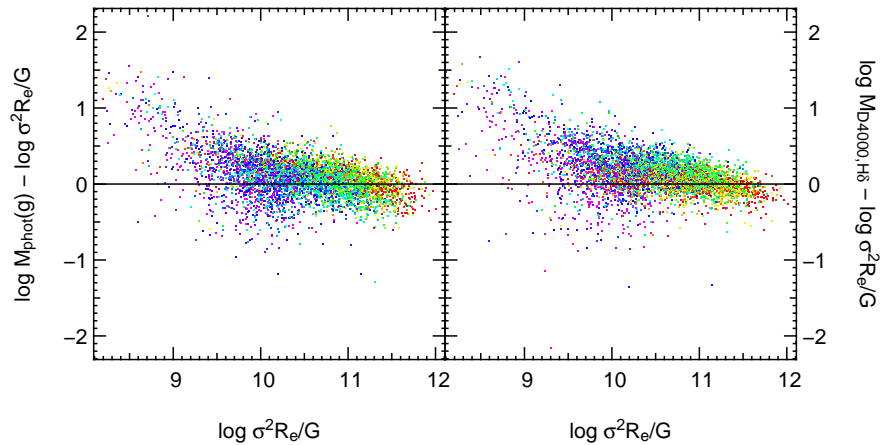}
  \caption{\label{fig:mass-dyn}%
    Comparison of stellar mass vs.\ dynamical mass, $\Mdyn \propto
    \sigma^2 R_e / G$. The left hand panel shows \Mdyn\ vs.\ 
    photometric mass and the right hand panel shows \Mdyn\ vs.\ the
    masses of K03. Color encodes \Ha\ equivalent width, as in Fig.~2.}
\end{figure*}

Since we do not have a fiducial mass estimator (neither for stellar
mass nor for total mass, for that matter), it is only natural to ask
how the stellar mass measurements presented here compare with
estimators of mass based on kinematic data. In fact, it has been
suggested that stellar mass (or, more accurately, baryonic mass) and
total mass are tightly related and that stellar mass can be used as a
surrogate for total mass in the context of high-$z$ galaxy surveys to
probe structure formation \citep[see, e.g.,][]{BE00}.

We use the measurements of velocity dispersion, $\sigma$, and
effective radius in the g band, $R_e$, provided by the SDSS pipeline
to plot stellar mass vs.\ dynamical mass, $\Mdyn \propto \sigma^2 R_e
/ G$, in Fig.~\ref{fig:mass-dyn}. The left hand panel shows \Mdyn\ 
vs.\ photometric stellar mass and the right hand panel shows \Mdyn\ 
vs.\ the stellar masses of K03. Color again encodes \Ha\ equivalent
width.

Above $10^{10}$~\Msun, the stellar masses from both methods follow the
dynamical masses remarkably well. Below $\sim 10^{9}$~\Msun, the
velocity dispersion measurements of the SDSS becomes unreliable as we
approach the instrumental resolution of the data ($\sim 70$~km
s$^{-1}$).

At higher masses, although both estimators generally follow \Mdyn,
there are again some differences, and both estimators show similar
trends in their residuals although with different amplitudes. Stellar
masses agree very well with \Mdyn\ at the highest masses (which are
mostly populated by old, quiescent objects). At lower masses, stellar
masses show a trend to larger values than \Mdyn\ with decreasing mass
and with increasing \Ha\ emission line equivalent width.  This effect
is weak in the photometric estimator, and stronger in K03's method,
which gives stellar masses larger than \Mdyn\ by 0.1 to 0.4~dex at
almost all masses. This comparison is unchanged by using \MLK\ instead
of \MLg.

It is important to note that \Mdyn\ is not a good estimator of total
mass, and that this comparison is again only to be taken in relative
terms. In fact $\sigma^2 R_e / G$ can only provide a lower limit to
the mass.  However, as long as a bulge is present, the total mass
should not be underestimated by more than $\sim 0.3$~dex (see, e.g.\ 
Fig.~4 in \citealp{WK81}; also \citealp{Padmanetal04}, who show that
$\sigma^2 R_e / G$ is a reasonable mass estimator, although this paper
is concerned with ellipticals only).  It is therefore not surprising
to find stellar masses in excess of \Mdyn\ and the difference between
the two increasing at lower masses.

Nevertheless, the point of this work is to assess the general
consistency and reliability of stellar mass estimates than to
investigate the relationship between stellar and dynamical mass in
galaxies. Fig.~\ref{fig:mass-dyn} shows that the estimators of stellar
mass, and especially the photometric estimator which is most easily
obtainable for large high redshift samples (covering a bluer
wavelength range, though), closely follow \Mdyn\ as measured by this
simple dynamical measure. We cannot see significant systematic
deviations which would bias or invalidate this estimator. This is a
very encouraging result, since such an estimator is very likely to be
the only one available at $z > 0$ for some time to come.

%% ------------------------------------------------------------------
%% ACKNOWLEDGMENTS
%% ------------------------------------------------------------------

\acknowledgments

This publication makes use of data products from the Two Micron All
Sky Survey, a joint project of the University of Massachusetts and the
Infrared Processing and Analysis Center/California Institute of
Technology, funded by NASA and NSF.  Funding for the creation and
distribution of the SDSS Archive has been provided by the Alfred P.
Sloan Foundation, the Participating Institutions, NASA, NSF, the U.S.
Department of Energy, the Japanese Monbukagakusho, and the Max Planck
Society.

N.D.\ acknowledges support by the Alexander von Humboldt Foundation.

%% ------------------------------------------------------------------
%% REFERENCES
%% ------------------------------------------------------------------

%\bibliography{apjmnemonic,literature} \bibliographystyle{apj}

\onecolumn

\end{document}